\documentstyle[11pt,twoside,jltp]{article}
\input{psfig}

\def\EJ{E_{\rm J}}
\def\CJ{C_{\rm J}}
\def\Cg{C_{\rm g}}
\def\Vg{V_{\rm x}}
\def\EC{E_{\rm C}}
\def\Cqb{C_{\rm qb}}
\def\Phix{\Phi_{\rm x}}
\def\Vtr{V_{\rm tr}}
\newcommand{\ket}[1]{\left|#1\right\rangle}
\newcommand{\Journal}[1]{{\it #1}}

\title{NANO-ELECTRONIC REALIZATIONS OF QUANTUM BITS}

\author{Yuriy Makhlin$^{\ast,\dagger}$\address{$^\ast$Institut f\"ur Theoretische
Festk\"orperphysik, Universit\"at Karlsruhe\\ 
D-76128 Karlsruhe, Germany\\
$^\dagger$Landau Institute for 
Theoretical Physics, Kosygin St. 2, 117940, Moscow, Russia\\
$^\ddagger$Forschungszentrum Karlsruhe, Institut f\"ur Nanotechnologie,
D-76021 Karlsruhe},
Gerd Sch\"on$^{\ast,\ddagger}$,
and Alexander Shnirman$^\ast$}

\runninghead{Yu.~Makhlin, G.~Sch\"on, and A.~Shnirman}{NANO-ELECTRONIC 
REALIZATIONS OF QUANTUM BITS}
       
\begin{document}

\begin{abstract}
Quantum computers could perform certain tasks which no classical computer can
perform in acceptable times.  Josephson junction circuits can serve as
building blocks 
of quantum computers.  We discuss and compare two designs, which employ charge
or magnetic flux degrees of freedom to process quantum information.  In both
cases, elementary single-qubit and two-qubit logic gates can be performed by
voltage or flux pulses.  The coherence time is long enough to allow a series of
such operations.  We also discuss the read-out,
i.e.\ a quantum measurement process.
In the charge case it is accomplished by coupling a single-electron 
transistor to the qubit.\\

\end{abstract}

\maketitle

\section{INTRODUCTION}
Quantum computers, if available, could perform certain operations in a
massive parallel way, which would lead to an exponential speed-up as
compared to the performance of classical computers.
Furthermore, medium-size quantum information processing devices could be
used in a number of applications such as, for instance, quantum
communication and quantum cryptography.  Several physical realizations of
quantum bits, i.e. the elementary building blocks of quantum
computers, have been 
proposed.  The best studied ones are trapped ions, NMR in the liquid state
and cavity QED.  Here we discuss solid-state nano-electronic
realizations.  They appear promising since they can be scaled
up to large numbers of qubits and are most easily embedded in
electronic circuits.   

Josephson junction circuits are particularly suitable  
for quantum information processing, since they combine the
intrinsic coherence of the superconducting state and the possibility
to control the circuit dynamics by voltage and magnetic flux pulses.  
Their fabrication and manipulation are possible 
by present-day technologies. The dynamical variables in these circuits
are the charges on the islands and the  phases of the superconducting
order parameter. Both are canonically conjugated, and  Heisenberg's uncertainty
principle holds for them. Depending on the
ratio of two characteristic energies -- the typical charging energy, which favors
well-defined charges, and the Josephson energy, which favors the phase degree of
freedom --  the charge or the flux can have a well-defined value, while the
other fluctuates strongly. Either charge~\cite{OurPRL,OurNature,Averin}
or phase~\cite{Feigelman,Mooij} degrees of 
freedom can be used to store and process quantum information.
Here we describe charge and phase (flux)
nano-electronic quantum bits and possible designs of the
circuits in both cases.

The basic elements of a quantum computer are the qubits, i.e.\ 
two-state quantum systems which can be manipulated, separately for each qubit,
by the control of the Hamiltonian.  This allows performing 
single-bit logic gates.  Additionally, one needs to be able to couple qubits in
a controlled way to perform two-bit logic operations.  The whole system should stay
coherent for a sufficiently long time, that is, the coupling to the environment
needs to be weak.  Finally, to read-out the information about the quantum state
of the system a quantum measurement should be performed.  Below we discuss these
steps for the relevant designs.

\section{JOSEPHSON-JUNCTION QUANTUM BITS}

\subsection{Charge qubit}

\begin{figure}
\centerline{\hbox{\psfig{figure=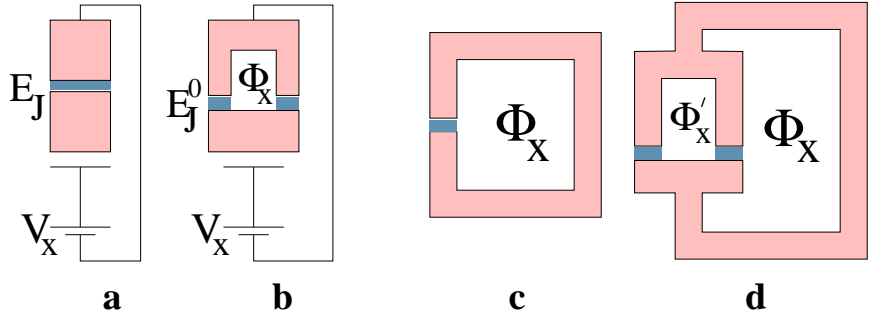,width=0.9\textwidth}}}
\caption[]{\label{Fig:Bits}%
{\bf a}. The simplest design of a charge qubit.
{\bf b}. A charge qubit with the flux-controlled Josephson coupling.
{\bf c}. The rf-SQUID, the simplest flux qubit.
{\bf d}. A flux qubit with controlled barrier height between the wells.
}
\end{figure}

The simplest design of the Josephson qubit using the charge degree of
freedom is presented in Fig.~\ref{Fig:Bits}a.  It consists of a
superconducting electron box with a
low-capacitance Josephson junction, with capacitance $\CJ$ and 
Josephson energy $\EJ$, biased by a voltage source through a gate
capacitor $\Cg$. 
If the superconducting gap is large enough then at low temperatures odd-parity
states are forbidden~\cite{Tinkham}, and the
charge on the superconducting island
is a multiple of the Cooper-pair charge $2ne$, measured relative to
the neutral state.  The 
Hamiltonian of the system is then given by
\begin{equation}
{\cal H}=\frac{(2ne-\Cg\Vg)^2}{2(\Cg+\CJ)}
-\EJ\cos\varphi \;,
\end{equation}
where the phase difference across the junction, $\varphi$, is canonically
conjugated to $n$.  The charging energy, with scale $\EC=e^2/2(\Cg+\CJ)$,
is chosen to dominate over the Josephson term $\EJ$.  In equilibrium at low
temperature, $k_{\rm B}T\ll\EC$, the 
system is in the ground state, which 
-- away from certain degeneracy points -- is 
approximately the charge state with the minimal charging
energy.  Only near the voltages $V_{\rm deg}=(2n+1)e/\Cg$, where the
states with $n$ 
and $n+1$ Cooper pairs on the island are degenerate, the weak Josephson
coupling mixes the charge states strongly.  Biased near  these voltages the
system reduces to a two-state problem, with the Hamiltonian in the
basis of charge states $\ket{n}$ and $\ket{n+1}$ 
\begin{equation}
{\cal H}=\frac{\varepsilon}{2} \hat\sigma_z
+\frac{\Delta}{2} \hat\sigma_x \;.
\label{HamEpsDelta}
\end{equation}
Here $\varepsilon(\Vg)=2e\frac{\Cqb}{\CJ}(\Vg-V_{\rm deg})$ denotes the 
difference in charging energy between two relevant charge states,
the tunneling amplitude between the states is $\Delta=-\EJ$, and the 
capacitance of the qubit in the circuit is $\Cqb^{-1}=\CJ^{-1}+\Cg^{-1}$.

The first term in the Hamiltonian (\ref{HamEpsDelta}) can be controlled through
the gate voltage.  The Josephson coupling in the second term 
can also be controlled if
the junction is replaced by the dc-SQUID threaded by a magnetic flux
$\Phix$, as shown in 
Fig.~\ref{Fig:Bits}b.  Then the effective Josephson energy is
$\EJ(\Phix)=2\EJ^0\cos(2\pi\Phix/\Phi_0)$.  With the control over
these two parameters one can perform any single-bit logic operation.
In the idle state between 
the operations one keeps the qubit at the degeneracy point $\Vg=V_{\rm deg}$ and
chooses $\Phix=\Phi_0/2$. Hence the Hamiltonian vanishes, 
${\cal H}=0$, and  the qubit's 
state does not evolve in time.  To perform an operation, one can
switch the flux to a different value 
for a finite time $\tau$.  The resulting change of the quantum state of
the qubit is described by the unitary operator $\exp(i\hat\sigma_x
\;\EJ\tau/2)$.  A voltage pulse 
results in another elementary operation, $\exp(-i\hat\sigma_z
\;\varepsilon(\Vg)\tau/2)$.  With a series of (no more than three) such
operations with proper time spans any $2\times 2$ unitary operation, that is any
single-qubit logic gate can be performed.  The typical times of the operations
are of order $\hbar/\Delta$ or $\hbar/\varepsilon$.

The manipulations of the state of the qubit are similar to the
manipulations used in NMR experiments, and various techniques familiar
from NMR applications can be employed.
For instance, instead of rectangular voltage pulses one can use
resonant ac-pulses to induce coherent transitions between the qubit's
states.  Then 
the typical operation time is determined by the amplitude of the ac-pulse and
can be optimized.  It should be slow enough to make the control easy
and fast enough to maximize the number of logic operations performed within the
phase coherence time.

\subsection{Flux qubit}
A controllable two-state quantum system can be realized also in the opposite
limit of dominating Josephson coupling.  The simplest design is the rf-SQUID 
(see
Fig.~\ref{Fig:Bits}c) that is a superconducting loop interrupted by a Josephson
junction.  The phase difference across the junction,
$2\pi\Phi/\Phi_0$, is controlled by the flux $\Phi$ in the loop, which
fluctuates  around the externally applied value $\Phix$.  With the
Josephson, charging and 
magnetic contributions taken into account, the Hamiltonian of the system reads
\begin{equation}
{\cal H}=-\EJ\cos\left(2\pi\frac{\Phi}{\Phi_0}\right)
+\frac{(\Phi-\Phix)^2}{2L} +\frac{Q^2}{2\CJ} \;.
\end{equation}
Here $L$ is the self-inductance of the loop and $\CJ$ the capacitance of the
junction.  The charge $Q=-(i/\hbar)\partial/\partial\Phi$ on the leads is
canonically conjugated to the flux $\Phi$.  If the self-inductance is large
($\beta_L\equiv \EJ /(\Phi_0^2/4\pi^2 L)$ is slightly larger than 1) and the
externally applied flux is close to $\Phi_0/2$, the two first terms in the
Hamiltonian form a double-well potential near $\Phi=\Phi_0/2$.  At low
temperatures only the lowest states in the two wells contribute to the physics
of the system.  The reduced Hamiltonian of this two-state system is 
again given by
Eq.(\ref{HamEpsDelta}), where now $\varepsilon(\Phix)= 4\pi\sqrt{6(\beta_L-1)}
\EJ(\Phix/\Phi_0-1/2)$ is the asymmetry of the double well potential,
and $\Delta$ is
the tunneling amplitude between the wells.  
The latter can be controlled through
the height of the barrier, which is determined by $\EJ$.  This
Josephson energy can be controlled, in turn, if 
the junction can be replaced by the dc-SQUID, as shown in
Fig.~\ref{Fig:Bits}d.

With two external parameters governing the Hamiltonian, elementary $z$- and
$x$-rotations can be performed, as we have seen in the previous subsection.
They can be driven either by switching the external flux for a finite time or by
resonant pulses.

The rf-SQUID described above was discussed in connection
with the `macroscopic quantum coherence', that is coherent
oscillations of a quantum system between two 
macroscopically different states \cite{LeggettMQC,Tesche}.  However, the
requirements of sufficiently large self-inductance and Josephson energy of the
junction make the rf-SQUID very susceptible to external noise, and the
experiments with the rf-SQUID were not successful so far.  To overcome this
difficulty Mooij et al.~\cite{Mooij} suggested to use a smaller superconducting
loop with three junctions (one with controllable critical current).  Then the
double-well potential is formed by Josephson terms for the junctions, and lower
critical currents can be used.  As a result the system should stay coherent for
a longer time.  We shall discuss dephasing effects in
Section~\ref{Section:Dephasing}.

\section{COUPLING OF THE QUBITS}
To perform a quantum computation with a register of qubits two-bit logic
operations are necessary.  For such an operation, which is a unitary
transformation of the quantum state of two qubits, the couplings between the
qubits should be controlled individually for each pair.  In this section we
discuss realization of such interactions for charge and flux qubits.

\subsection{Controlled coupling of charge qubits}
\begin{figure}
\centerline{\hbox{\psfig{figure=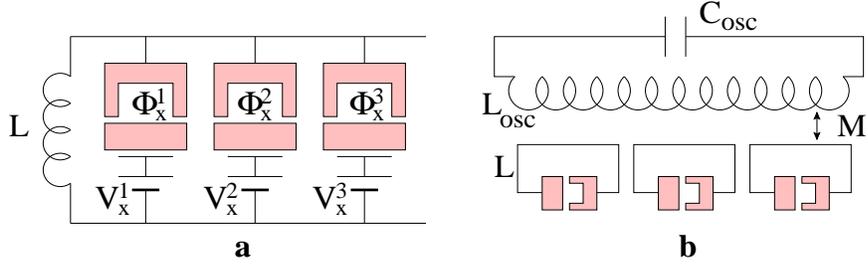,width=0.9\textwidth}}}
\caption[]{\label{Fig:CoupledBits}%
Coupling of qubits via an oscillator circuit: {\bf a}. for charge qubits and 
{\bf b}. for flux qubits.
}
\end{figure}

One possibility to couple charge qubits is to join them in
parallel to a common inductor, as pictured in Fig.~\ref{Fig:CoupledBits}a.  Then
their dynamics is coupled to the oscillations in the
$LC$-circuit, which is formed by the inductor and the capacitances of all $N$ 
qubits in
parallel.  If the frequency $\omega_{LC}=1/\sqrt{LN\Cqb}$ of these oscillations
is higher than typical qubit frequencies, the oscillatory degrees of
freedom are not excited by the qubit manipulations, but still they provide the
effective coupling between the qubits~\cite{OurPRL,OurBook}.

To clarify the physics of the coupling and to estimate its magnitude, we provide
a simple derivation.  The current through the inductor, $I$, is given by
the contributions of all qubits, $I=\sum_i I^i$.  In
terms of the flux $\Phi$ through the inductor the Hamiltonian of the
oscillations is expressed as ${\cal H}_{\rm
osc}=\frac{\Phi^2}{2L}-I\Phi+\frac{Q^2}{2N\Cqb}$ or
\begin{equation}
{\cal H}_{\rm osc}=\frac{(\Phi-LI)^2}{2L}+\frac{Q^2}{2N\Cqb}-\frac{LI^2}{2} 
\;.
\label{HamLC}
\end{equation}
Here $Q=-(i/\hbar)\partial/\partial\Phi$ is the charge canonically conjugated to
$\Phi$.  If the time-evolution of the qubits is slow compared to the
oscillations, we can use an adiabatic approximation and consider $I$ as
constant, and the oscillator  remains in
its ground state.  The energy of the ground state of the first two terms in
(\ref{HamLC}), $\hbar\omega_{LC}/2$, does not depend on $I$, while the last term
provides the current-dependent correction.  This correction describes the
effective coupling between the qubits, $-L(\sum_i I_i)^2/2$.  In the basis of
the qubits' charge states individual current operators can be expressed as
$I^i=(\Cqb/\CJ)(e\EJ^i/\hbar) \hat\sigma_y^i$, leading to the interaction term
\begin{equation}
{\cal H}_{\rm int}=-\pi^2\left(\frac{\Cqb}{\CJ}\right)^2
\sum\limits_{i<j}\frac{\EJ(\Phix^i)\EJ(\Phix^j)}{\Phi_0^2/L}\;
\hat\sigma_y^i \hat\sigma_y^j
\; .
\end{equation}
This coupling can be controlled through external fluxes $\Phix^i$, which bias
the SQUIDs of the qubits.  Between the operations we keep all
$\Phix^i=\Phi_0/2$, so that $\EJ^i=0$ and the coupling is off.  During
single-bit operations we switch on $\EJ^i$ for at most one qubit, so that the
interaction term is still zero.  When a two-qubit operation is needed, the
fluxes are changed for two qubits, providing the interaction between them for a
finite period.  This interaction leads to a non-trivial unitary transformation
of the quantum state of two chosen quantum bits, i.e.  to a two-bit logic gate.

This design allows performing both single- and two-bit operations by
controlling  the same
external parameters (fluxes and voltages) as for individual qubits.

\subsection{Coupling of flux qubits}

One possibility to couple flux qubits is to use a flux transformer~\cite{Mooij}
which provides an inductive coupling between the qubits.  Each qubit has two
loops, those of the rf- and dc-SQUIDs in the simplest rf-SQUID design.  Any loop
of one qubit can be coupled to any loop of the other, giving rise to
$\hat\sigma_z^i \hat\sigma_z^j$, $\hat\sigma_z^i \hat\sigma_x^j$ or
$\hat\sigma_x^i \hat\sigma_x^j$ coupling terms.  To turn off this coupling
completely, one would need to have an ideal switch in the flux transformer.
This switch is to be controlled by high-frequency pulses, and the related
external circuit can lead to decoherence effects.  An alternative is
to keep the interaction on constantly and use ac driving pulses to induce
coherent transitions between the levels of the two-qubit system
(cf.~Refs.~\citen{OurPRL,Mooij}).  A disadvantage of this approach is that
permanent couplings lead to unwanted 2-qubit correlations between the logic
operations.

Here we discuss another design of a many-qubit circuit, which allows to control
interactions via flux sources of individual bits.  The circuit, shown in
Fig.~\ref{Fig:CoupledBits}b, is similar to the register of charge qubits
discussed in the previous subsection.  It includes an $LC$-circuit formed by a
loop, with the self-inductance $L_{\rm osc}$, interrupted by a small capacitor
$C_{\rm osc}$.  The loop is coupled inductively to the set of flux qubits and
mediates interaction between them.  Following the derivation in
Refs.~\citen{OurPRL,OurBook} one can find the effective coupling by integrating out
the oscillations which are faster than the qubit's dynamics.  A simple way to
obtain the coupling is to notice that in the low-capacitance limit, $C_{\rm
osc}\to 0$ (almost decoupled qubits) the effect of the qubits is to establish a
voltage drop across the inductor $L_{\rm osc}$, given by $V=\sum_i
M\dot\Phi_i/L$.  Here for each qubit $\Phi_i$ is the flux in its loop, $L$ is 
the
self-inductance of the loop and $M$ is the mutual inductance with the loop of
the $LC$-circuit.  Then the Hamiltonian for the oscillator mode is ${\cal
H}_{\rm osc}=\Phi^2/2L_{\rm osc} + Q^2/2C_{\rm osc} - VQ$, with
$Q$ being the charge conjugated to the flux $\Phi$ through the $LC$-circuit.
By similar arguments as in the previous subsection we find that the oscillator
provides the inter-qubit interaction term $-C_{\rm osc}V^2/2$.  In limit of the
weak coupling to the $LC$-circuit, the time derivatives of the qubits' fluxes are
given by $\dot\Phi_i=\frac{i}{\hbar}[{\cal H}_i,\Phi_i]=\delta\Phi_i\Delta_i
\hat\sigma_y^i/\hbar$ where $\delta\Phi_i$ is the separation between two minima
of the potential of the rf-SQUID.  Hence, the interaction is given by
\begin{equation}
{\cal H}_{\rm int}=-\pi^2
\left(\frac{M}{L}\right)^2
\sum\limits_{i<j} 
\frac{\delta\Phi_i\delta\Phi_j}{\Phi_0^2}
\frac{\Delta_i\Delta_j}{e^2/C_{\rm osc}}\;
\hat\sigma_y^i\hat\sigma_y^j \;.
\end{equation}
This interaction can be controlled through tunneling amplitudes $\Delta_i$ of
individual qubits.  To turn off the interaction, one should 
put $\Delta_i$ to zero. The amplitudes can be exponentially suppressed by
increasing the barrier heights. (Note that in this case also
the fluctuations of  $\Delta_i$ are exponentially
suppressed, unlike for the charge qubits, where stronger fluctuations
are present.)  The absence of the interaction
between the operations helps increasing the accuracy of the manipulations.

\section{ENVIRONMENT AND DEPHASING}
\label{Section:Dephasing}

Due to the unavoidable coupling to environmental degrees of freedom
the quantum state 
of the qubits gets entangled with the environment, which leads to dephasing
effects.  Charge qubits are sensitive to the electromagnetic
fluctuations in the external circuit and in the substrate and to
background charge fluctuations.  Flux qubits are
insensitive to the latter.  Various sources of
dissipation for flux qubits are discussed in Ref.~\citen{Mooij}.  Here we
estimate the effect of fluctuations in the external circuit.

The control over the voltage or flux requires 
coupling the quantum system  
to a dissipative external circuit, which introduces fluctuation effects.
The voltage fluctuations in charge qubits or the flux fluctuations in the
main loop of  
flux qubits are coupled to $\sigma_z$-degree of freedom.
The external circuit can be parameterized
by an effective impedance of the voltage (flux) source, which in turn can be
modeled by an oscillator bath in the spirit of the Caldeira-Leggett
model. The behavior of  
a two-state system coupled to an oscillator bath has been discussed in
the context of the spin-boson model~\cite{Leggett,Weiss}. 
The conclusion is that
the decay of the off-diagonal elements of the density matrix of the qubit 
(dephasing) and the relaxation of the diagonal elements to their 
equilibrium values are described by the two time scales
\begin{eqnarray}
\tau_\varphi&=&\gamma\;\;\;
\frac{\hbar}{\Delta E}\;\;\;
\left(
\frac{1}{2}\coth\frac{\Delta E}{2k_{\rm B}T}\sin^2\eta+
\frac{2k_{\rm B}T}{\Delta E}\cos^2\eta
\right)^{-1}
\label{dephasing}
\; ,
\\
\tau_{\rm relax}&=&\gamma\;\;\;
\frac{\hbar}{\Delta E} \;\;\;
\left(\coth\frac{\Delta E}{2k_{\rm B}T}\sin^2\eta\right)^{-1}
\label{relaxation}
\; ,
\end{eqnarray}
respectively. Here $\hbar/\Delta E$ gives the typical operation time (cf.
Section 2) in terms of the level spacing $\Delta 
E=\sqrt{\varepsilon^2+\Delta^2}$ of the qubit, $T$ is the temperature, and 
$\tan\eta\equiv\Delta/\varepsilon$.

The dimensionless parameter $\gamma$ describes the strength of the dissipation.
For charge qubits it is determined by the resistance $R_V$ of the voltage 
source in units of  the large quantum resistance $R_{\rm K}=h/e^2 \approx
25.8 {\rm k}\Omega$,
\begin{equation}
\gamma_V=\frac{1}{4\pi}\frac{R_{\rm K}}{R_V}
\left(\frac{\CJ}{\Cqb}\right)^2
\; .
\label{gammaV}
\end{equation}
A small
gate capacitance $\Cg\approx\Cqb\ll\CJ$   provides a
weak coupling between qubit and environment.

For flux qubits, $\gamma$ is fixed by the impedance $R_I$ of the current source 
in the input loop, which provides the flux bias,
\begin{equation}
\gamma_I=
a
\frac{R_I}{R_{\rm K}}
\left(\frac{\Phi_0^2/4\pi^2M}{\EJ}\right)^2
\;.
\label{gammaI}
\end{equation}
Here $M$ is the mutual inductance of the input loop and the qubit's loop.
The numerical prefactor in Eq.\ (\ref{gammaI}) is 
$a=\pi/(6(\beta_L-1))$ for the rf-SQUID and is also of the order of unity for 
the design of Ref.~\citen{Mooij}.  The dephasing is slow for small loops
and junctions with low critical currents.  Indeed, the argument in the 
bracket in
(\ref{gammaI}) is proportional to $\Phi_0^2/4\pi^2L\EJ$.
While this quantity should be slightly smaller than one for the rf-SQUID,
it can be much larger for the design of Mooij et al.\cite{Mooij},
corresponding to slower dephasing.

The qubit is equivalent to a spin-$1/2$ particle in the external magnetic
field $\varepsilon\hat z+\Delta\hat x$, while the environment produces
 an additional
fluctuating field in $\hat z$-direction.  The component of the fluctuating field
orthogonal to the external field (with the magnitude proportional to $\sin\eta$)
induces transitions between the eigenlevels, while the longitudinal component
($\propto\cos\eta$) leads to random fluctuations of eigenenergies.  Only the
former process leads to relaxation (\ref{relaxation}), while both contribute to
the dephasing (\ref{dephasing}).  This observation explains the
$\eta$-dependence of 
the decoherence rates (\ref{dephasing}), (\ref{relaxation}).  The relaxation
rate is slow for low-impedance  voltage sources and high-impedance
current sources.  Weak coupling to the external circuit, i.e., small 
$\Cg$ or small $M$, helps further maintaining the coherence.

The effect of fluctuations of the $\sigma_x$-term in the Hamiltonian (in the
flux circuit of the dc-SQUID-loop which controls the Josephson coupling in both
charge and flux qubits) can be described in a similar way~\cite{OurNature}.
Because of the different direction of the fluctuating effective `magnetic
field', $\sin\eta$ and $\cos\eta$ should be interchanged in those terms.  
The effect of these terms is relatively weak for the operation regimes discussed 
in Refs.~\citen{OurNature} and \citen{Mooij}.

\section{READ-OUT: QUANTUM MEASUREMENT}

\begin{figure}
\centerline{\hbox{\psfig{figure=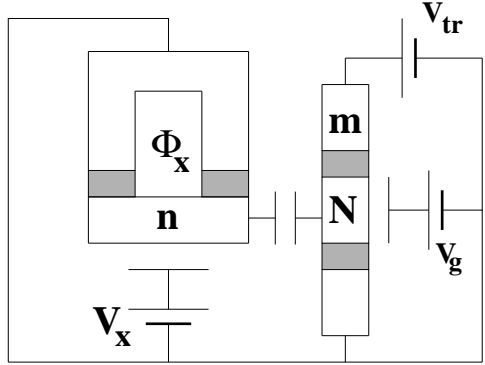,width=0.5\textwidth}}}
\caption[]{\label{Fig:Measurement}%
Read-out is accomplished by coupling a single-electron transistor capacitively 
to the qubit.
}
\end{figure}

To complete the quantum computation, one needs to read out the information about
the quantum state of the qubits, that is to perform a quantum measurement.  For
charge qubits this can be accomplished by coupling the qubit capacitively to a
single-electron transistor~\cite{OurPRB} as shown in Fig.~\ref{Fig:Measurement}.
During the computation the SET is kept in the off-state $\Vtr=0$, with no
dissipative current and no additional decoherence.  The only effect of the
transistor is a renormalization of the capacitances in the qubit's circuit.  To
perform the measurement, the transport voltage is switched to a sufficiently
high value and the dissipative current starts to flow.  The value of the current
in the circuit of the transistor is very sensitive to the charge on the island
of the qubit.  Monitoring the current, one can extract from the data information
about the state of the qubit.

To study this quantum measurement process, we analyzed the time evolution of the
density matrix of the coupled system of the qubit and the
SET~\cite{OurPRB,OurBook}.  The system is characterized by three energy scales:
the typical Coulomb energy of the transistor, $E_{\rm set}$, the charging energy
of the qubit, $\varepsilon(\Vg)$, and the Coulomb interaction between the
charges of the qubit and the middle island of the SET, $E_{\rm int}$.  We choose
$E_{\rm set}$ to be the largest energy scale, $E_{\rm set}\gg \varepsilon\gg
E_{\rm int}$.  The Josephson coupling $\EJ(\Phix)$ is switched to a small (or
zero) value before the measurement.  The transport voltage should be large
enough, of order $E_{\rm set}/e$, to overcome the Coulomb energy gap between two
charge states of the SET.  The value of this gap is slightly different for
different states of the qubit, which leads to different tunneling rates
$\Gamma\pm\delta \Gamma$ through the transistor, with $\Gamma$ of order
$2\pi\alpha E_{\rm set}$ and $\delta\Gamma$ of order $2\pi\alpha E_{\rm int}$.
Here $2\pi\alpha\equiv R_{\rm K}/4\pi^2 R_{\rm T}\ll 1$ is the dimensionless
tunneling conductance of the junctions of the transistor, of the
same order in both junctions. For definiteness we consider the limit $E_{\rm 
int}\ll \Gamma\ll \Delta E$, where essential features of the evolution can be 
seen, although the derivation can be performed in a wider parameter range.

To describe the system we single out the most important degrees of freedom:  the
qubit's degree of freedom, the charge $Ne$ on the middle island of the SET and
the charge $me$ which passed through the transistor after the
transport voltage is turned on, that is the time integral of the
current.  The microscopic degrees of 
freedom can be traced out, and a closed set of equations can be
derived~\cite{OurPRB,Schoeller} for the elements of the
density matrix, $\rho^i_j(N,m)$ which are diagonal in $N$ and $m$.

The evolution of the system is characterized by three time scales.  In the first
stage the random tunneling processes in the SET lead to the loss of the phase
coherence of the qubit.  The off-diagonal elements of the reduced $2\times2$
density matrix of the qubit, $\sum_{N,m}\rho^i_j(N,m)$, decay to zero on a short
time scale of order $\tau_\varphi\approx \Gamma/E_{\rm int}^2$.

At a later stage the information about the qubit's state can be deduced from the
current.  The dynamics of the current in the SET can be described by the
probability distribution of the number of electrons $m$ which have passed
through the transistor:  $P(m,t)\equiv\sum_{i,N}\rho^i_i(N,m)$.  At
the start of the measurement
$t=0$ no electrons have tunneled, and $P(m,t)=\delta_{m,0}$.  Then the
current begins to flow, and the peak shifts in the direction of positive $m$ and
widens due to shot noise effects.  Since the two charge states of the qubit
correspond to different net currents in the SET, after some time the peak splits
into two.  The separation of the peak centers grows linearly with time,
$2\delta\Gamma t$, while their width grows as $\sqrt{\Gamma t}$.  Therefore they
separate after time $t_{\rm meas}\approx \Gamma/\delta\Gamma^2$.
The weights of
the peaks after the separation are given by $|a|^2$ and $|b|^2$
for the initial state of the qubit $a\ket{0}+b\ket{1}$, and a good
quantum measurement is realized if $m$ is measured after $t_{\rm meas}$. As 
expected, the measurement time is longer than the dephasing time.

This ideal picture gets more complicated at even longer times.  At finite, or
fluctuating, $\EJ$ the eigenstates of the qubit are different for different
charges $Ne$ of the SET, and the measurement induces transitions
between them.  After a long mixing time $t_{\rm
mix}\approx \Delta E^4/(E_{\rm int}^2\EJ^2\Gamma)$ the diagonal elements of the 
density matrix of the qubit become equal
to $1/2$, and the information about their initial values is lost completely.  At
the same time $P(m,t)$ develops a plateau between the two peaks.  The weights of
the peaks decay exponentially.  At longer times $t\gg t_{\rm mix}$ the plateau
transforms into a peak around $m=\Gamma t$.  This distribution  contains
no information about the initial state of the qubit anymore.

\begin{figure}
\vskip5mm
\centerline{\hbox{\psfig{figure=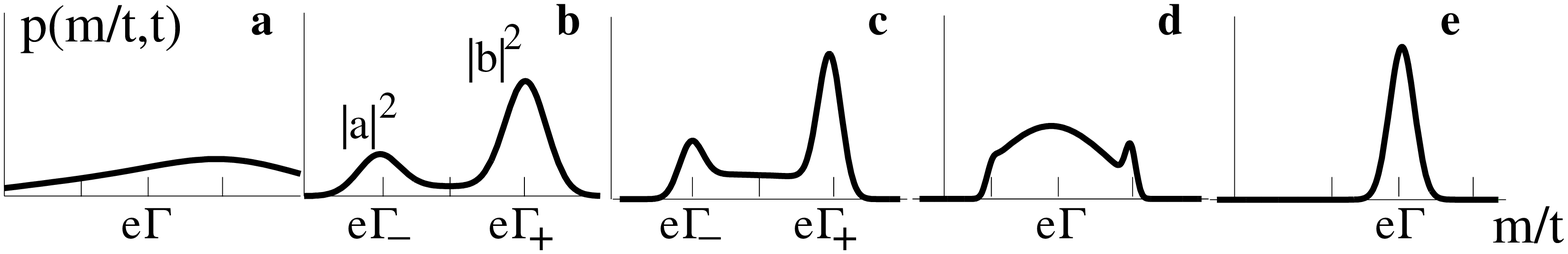,width=\textwidth}}}
\caption[]{\label{Fig:Pmt}%
The evolution of the probability distribution $p(m/t,t)$ for the values of the 
average current in the SET. The initial state of the qubit is
$a=\sqrt{1/4}$, $b=\sqrt{3/4}$. {\bf a}. $t<t_{\rm meas}$:
a very broad peak due 
to shot noise. {\bf b}. $t_{\rm meas}<t<t_{\rm mix}$: two peaks separate.
{\bf c--e}. $t\sim t_{\rm mix}$ 
and $t>t_{\rm mix}$: a plateau grows between the two peaks, erasing the 
information about the initial state of the qubit.
}
\end{figure}

The time evolution of $P(m,t)$ is depicted in Fig.~\ref{Fig:Pmt}.  For 
convenience the probability distribution $p(m/t,t)$ of possible values of the
quantity $m/t$, the current in the SET averaged over time $t$, is shown.  It is
related to $P(m,t)$ by $p(\bar I,t)=t\;P(\bar I t,t)$ where the prefactor $t$
ensures the normalization of the distribution.  If the current is averaged over
short times $t<t_{\rm meas}$, Fig.~\ref{Fig:Pmt}a, the shot noise does not allow
to distinguish between close values of the current corresponding to different
qubit states (very broad peak around $e\Gamma$).  At longer times $t_{\rm
meas}<t<t_{\rm mix}$, Fig.~\ref{Fig:Pmt}b, two peaks are formed around
$e\Gamma_\pm\equiv e(\Gamma\pm\delta\Gamma)$, and the measurement can be performed.  At $t\sim
t_{\rm mix}$ a plateau starts to grow between the peaks, and at the very long
times $t\gg t_{\rm mix}$ the plateau takes over and transforms into a narrow
peak around $e\Gamma$.

\section{CONCLUSION}

We have discussed two possible designs of nano-electronic quantum
bits, based on
Josephson junction circuits in the charge and flux regime.  In both
cases the
quantum dynamics of the qubits can be controlled through voltages and fluxes
(currents).  We suggest a way to couple flux qubits which is a dual analog of
the approach suggested earlier for charge qubits.  Compared to the proposal of
Mooij et al.~\cite{Mooij} it has the advantage that it allows to control the
inter-qubit interaction without introducing new links between the qubits and the
external circuit.  Apart from that we estimated the dephasing times due to
coupling to electromagnetic fluctuations in the external circuit.  This
contribution has not been discussed earlier for flux qubits.  The
quantum measurement of the state was discussed in detail for charge qubits.
We provided an analysis of the long-time evolution of the probability
distribution of 
current values in the single-electron transistor and of the qubit's density
matrix.  The design of a quantum measurement circuit for flux qubits requires
further investigation~\cite{Mooij}.

Earlier experiments with the superconducting box have demonstrated the quantum
nature of the two-state system and superpositions of different charge
states~\cite{Bouchiat,Nakamura}.  Recently Nakamura et al.~\cite{NakamuraNature}
observed time-resolved coherent oscillations of the quantum state of a charge
qubit.  Their experiment is the first observation of the macroscopic quantum
coherence (MQC) effect and realizes a single-bit operation (corresponding to an
$x$-rotation).  The observed phase coherence time of several nanoseconds was
shorter than the value $\sim 100$~ns, predicted by (\ref{dephasing}),
(\ref{relaxation}), (\ref{gammaV}) because of the use of a simple (but
efficient) measuring procedure.  Further development of the measurement
apparatus (cf.  Ref.~\citen{RFSET}) can render the coherence time longer.  In
flux systems, the attempts to observe MQC with the rf-SQUID were not successful,
but recent developments~\cite{Mooij} should make it possible in the near future.

\end{document}